\documentclass[preprint]{rsl}
\usepackage{url}
\usepackage{titlesec}
\usepackage{hyperref}
\usepackage{csquotes}
\titlespacing{\section}{0pt}{5pt}{1pt}
\titlespacing{\subsection}{0pt}{5pt}{1pt}
\titlespacing{\subsubsection}{0pt}{5pt}{1pt}
\title{Discovering Millisecond Pulsars in Globular Clusters with the GMRT (GCGPS)}
\author{Jyotirmoy Das, Jayanta Roy, Paulo C. C. Freire, Scott M. Ransom, Bhaswati Bhattacharyya, Karel Adamek, Wesly Armour, Sanjay Kudale, Mekhala Muley}

\begin{document}
\maketitle


\begin{abstract}
  The Globular Clusters GMRT Pulsar Search (GCGPS) project, launched in mid-May 2023, has emerged as one of the most successful pulsar surveys conducted with the upgraded Giant Metrewave Radio Telescope (uGMRT), leading to the discovery of several new millisecond pulsars (MSPs). The recently concluded Phase I of the survey resulted in the discovery of seven MSPs across four globular clusters (GCs), all of which previously had no known pulsars. These discoveries have enabled the precise determination of the dispersion measures (DMs) for the respective clusters for the first time. Consistent timing follow-up revealed that out of the seven MSPs, three are confirmed binaries, while two are isolated. This paper summarises the design and the implementation of Phase I of the GCGPS project, presents the key scientific results obtained so far, and outlines the strategy and progress of the recently initiated Phase II observations.
\end{abstract}

\section{Introduction}
Globular clusters (GCs) are dense, gravitationally bound stellar systems containing tens of thousands to millions of stars. Their high stellar densities make them efficient formation sites for exotic pulsar systems, far more frequently than in the Galactic field. GC pulsars span a wide range, from slow pulsars to millisecond pulsars (MSPs), and from isolated objects to diverse binary systems, leading to numerous important scientific discoveries~\cite{Ridolfi_2019, Barr_2024}. To date, about 359 pulsars have been discovered in 46 GCs, over 210 of which are confirmed binaries. The dense dynamical environments of GCs drive the formation and evolution of interacting systems such as black widows, redbacks, and transitional MSPs, which are expected to be more common in clusters than in the Galactic field. GCs also host some of the most extreme systems known, including the fastest-spinning pulsar~\cite{Hessels_2006}, highly eccentric binaries~\cite{Freire_2004}, and double neutron star systems~\cite{Tauris_2017}. They may even contain ultra-fast spinning pulsars, such as sub-millisecond MSPs, which would place strong constraints on the neutron star equation of state~\cite{Lattimer_2001}. Furthermore, cluster cores could harbour rare systems such as MSP-black hole or MSP-MSP binaries, offering unique opportunities to study pulsar evolution and to test theories of gravity~\cite{Kramer_2021}.

\section{Motivation}
The motivation for designing a survey to search for MSPs in globular clusters using the upgraded Giant Metrewave Radio Telescope (uGMRT) arose from its unique low-frequency capabilities and array
configuration. The uGMRT’s wide low-frequency coverage ($\sim$120 to 1460 MHz), combined with its Y-shaped array geometry comprising a total of 30 antennas (about 14 centrally clustered and 16 in the three arms), enables the formation of target-specific phased-array beams, making it highly efficient for deep pulsar searches in crowded GC fields. In contrast, most earlier surveys conducted with telescopes such as the GBT, Parkes, and Arecibo were primarily carried out at GHz frequencies, typically in L-band, and in some cases S-band for the GBT~\cite{Ransom_2004,  Hessels_2007}. As a result, these surveys were inherently biased toward detecting flat-spectrum MSPs. The uGMRT, with its substantially improved sensitivity at low frequencies, offers a powerful opportunity to probe steep-spectrum MSPs that previous high-frequency surveys may have missed. Furthermore, a large number of GCs remain either unexplored or only sparsely observed for pulsars, providing strong additional motivation for initiating the Globular Cluster GMRT Pulsar Search (GCGPS) using the uGMRT.

\section{The Globular Clusters GMRT Pulsar Search (GCGPS) survey}
The GCGPS survey has two distinct phases in its operational timeline, each with a different strategy and execution plan. In this section, we discuss both phases separately, along with their corresponding scientific outcomes.

\subsection{GCGPS PHASE I}
We have been conducting GCGPS Phase I since mid-2023, with the primary goal of discovering faint MSPs, which typically exhibit steep radio spectra and are therefore best probed at low frequencies (as uGMRT Band-3 and Band-4 with central frequencies, $\rm f_\mathrm{c} \sim$ a few hundred MHz). The survey design, target selection, sensitivity analysis, and the results from GCGPS Phase I are described below.

\subsubsection{Designing the survey}
\label{survey_design}
In designing the initial strategy for the GCGPS survey (Phase I), we adopted a phased-array (PA) beam configuration on the target globular cluster instead of an incoherent-array (IA) beam, prioritising higher sensitivity over wider sky coverage. To determine the optimal PA beam setup, we simulated uGMRT PA beams in both Band-3 (300 MHz - 500 MHz) and Band-4 (550 MHz - 750 MHz) using different antenna combinations to explore the trade-off between sensitivity and field-of-view. These simulations showed that a Band-4 PA beam formed using the 14 closely spaced antennas of the Central Square (CSQ) provides the best balance between sensitivity and sky
coverage, enabling full coverage of most target clusters. In addition to the CSQ beam, a second PA beam was formed using antennas up to the third antenna on each arm (\enquote{3rd Arm} beam), comprising 22 antennas. This configuration yields a narrower but more sensitive beam toward the cluster core, where most pulsars are expected. Accordingly, each cluster in GCGPS Phase I was observed simultaneously with two PA beams, the CSQ and the 3rd Arm beam, using Band-4 as the primary band. Band-3 observations were used only for clusters too extended to be fully covered by the Band-4 beams.

\subsubsection{Survey sensitivity}
\label{survey_sensitivity}
For our GCGPS Phase I survey, the typical on-source time per GC was about 2.5 hours. Now, according to the Radiometer equation, we can write:
\begin{equation}
    \rm S_\mathrm{min} = \frac{SNR \times T_\mathrm{sys} \times \beta}{G \times N \times \sqrt{n_\mathrm{pol} \times BW \times \Delta t_\mathrm{obs}}} \sqrt{\frac{W_\mathrm{eff}}{P_\mathrm{s} - W_\mathrm{eff}}}
\end{equation}

Where $\rm W_\mathrm{eff}$ is the effective pulse width, including dispersion broadening and scattering delay. For the uGMRT, the per-antenna gain is $\rm G \sim 0.35~K Jy^{-1}$ with number of polarisation channels ($\rm n_\mathrm{pol}$) = 2. For Band-4 ($\rm BW=200$ MHz), the system temperature ($\rm T_\mathrm{sys}$) is about 102.5 K at 650 MHz. Assuming observations with the CSQ beam ($N=14$ antennas), an on-source integration time ($\Delta t_\mathrm{obs}$) of 2.5 hours, and negligible quantisation loss ($\beta\sim1$) for 8-bit digitisation, we estimate the theoretical $10\sigma$ detection sensitivity of the GCGPS Phase I survey. For a pulsar with a 10\% duty cycle, spin period of 4.60 ms (the median period of known GC MSPs), and $\rm DM=50~pc,cm^{-3}$, the resulting sensitivity is $\sim48~\mu$Jy. For comparison, MeerKAT's TRAPUM L-band ($\rm f_\mathrm{c} \sim 1280\:MHz$) 10$\sigma$ detection sensitivity is $\rm \sim\:10\:\mu Jy$ (Ridolfi et al. 2022~\cite{Ridolfi_2022}), having the same on-source time. Scaling this sensitivity to 650 MHz for steep spectrum sources (spectral index ~$\sim$\:-2), the sensitivity becomes comparable to that of GCGPS Phase I. Notably, TRAPUM is the most sensitive GC MSP search project carried out in the southern hemisphere. Now, for the 3rd Arm beam compared to our CSQ beam, the central sensitivity will scale by a factor of the ratio of antenna number to the CSQ beam, which is $\sim$ 1.57 (N2/N1 = 22/14), with a comparatively lower sky coverage than the CSQ beam.

\subsubsection{Target selection}
\label{target_selection}
Due to the superior sensitivity of the Five-hundred-meter Aperture Spherical Telescope (FAST,~\cite{Zhang_2023}), we excluded the FAST-accessible sky (declination $\geq$ -17 degrees) from the GCGPS survey, restricting the Phase I sky coverage to -53 $\leq\:\delta\:\leq\:-17$ degrees. Similarly, globular clusters already observed by the MeerKAT TRAPUM survey were not considered, given their comparable sensitivity. We therefore focused primarily on GCs previously observed in L-band surveys with the GBT and Parkes, where uGMRT Band-4 observations are expected to provide an improvement in sensitivity of about 3 to 4 times for steep-spectrum MSPs (spectral index $\sim$ -2). In addition, clusters that had never been searched for pulsars were also included. All candidate GCs were ranked according to their likelihood of hosting MSPs, using X-ray brightness, total cluster mass, and compactness as selection criteria. Based on this ranking, 31 of the most promising clusters were selected for observations under GCGPS Phase I across multiple uGMRT cycles. The full list of selected clusters is provided in Table 2 of Das et al. 2025~\cite{Das_2025}.

\begin{figure}
    \centering
    \includegraphics[width=\linewidth]{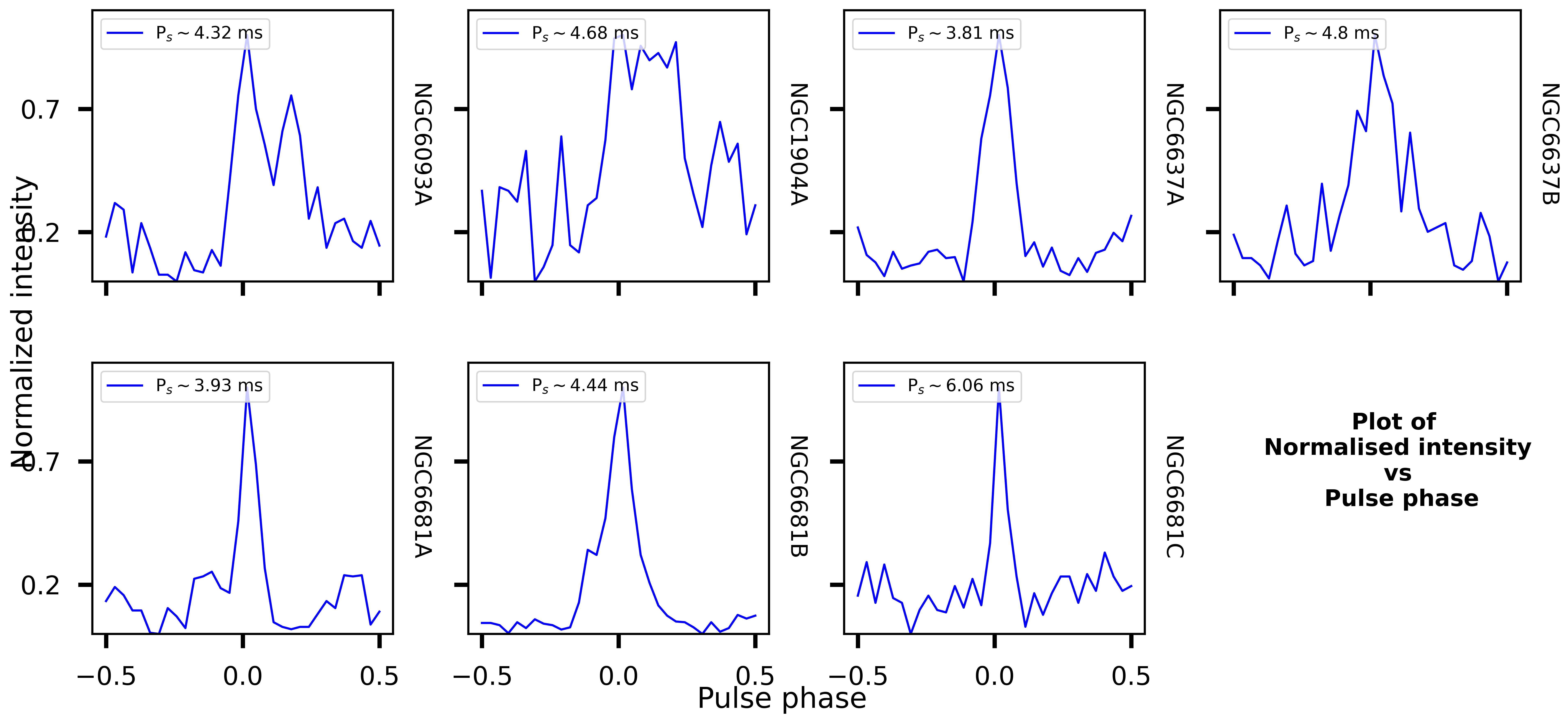}
    \caption{Normalised folded profiles of all seven MSPs discovered across four GCs under the GCGPS Phase I survey.}
    \label{fig:gcgps_msps}
\end{figure}

\subsubsection{Observation and data analysis}
We began GCGPS Phase I observations in May 2023, simultaneously observing each GC with two phased-array beams (the CSQ and the 3rd Arm beams; see Section~\ref{survey_design}). We record 8-bit data over a 200 MHz bandwidth with 4096 frequency channels and an 81.92~$\mu$s time sampling, using a typical on-source integration time of $\sim$2.5 hours (see Section~\ref{survey_sensitivity}) to enable a high-time-resolution search for GC MSPs. We also record simultaneous interferometric data (using all 30 dishes), facilitating the image plane localisation upon discovery. To process the obtained beam data, we developed an end-to-end {\tt PRESTO}-based acceleration-search pipeline, {\tt PSS}\footnote{{\tt PSS} GitHub repository: \url{https://github.com/jyotirmoydas5392/Pulsar_Search_Script.git}}. We perform FFT-based acceleration searches with $Z_{\max}=200$ up to eight harmonics, making the search sensitive to compact binary MSPs. The results we obtained till now with these GCGPS observations and data processing are described below.

\begin{figure}
    \centering
    \includegraphics[width=\linewidth]{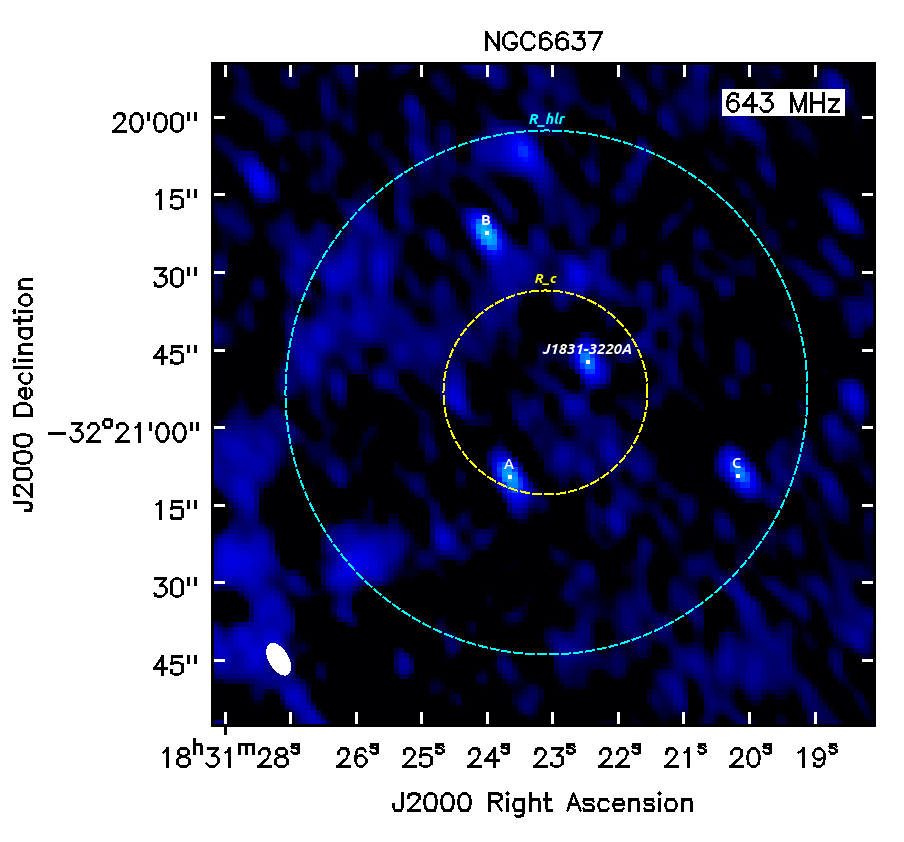}
    \caption{Localised position of NGC6637A (J1831$-$3220A) discovered in the GCGPS Phase I observation of NGC 6637 (M69). The inner yellow circle is the core radius, while the outer cyan circle is the half-light radius of M69.}
    \label{fig:NGC6637A_localisation}
\end{figure}

\subsubsection{Discoveries and timing results}
Over the course of a few years' observing campaign, we discovered a total of seven MSPs (see Figure~\ref{fig:gcgps_msps}) in four globular clusters that previously had no known pulsars. All discoveries from Phase I have been updated on the GCGPS website\footnote{GCGPS website: \url{https://www.ncra.tifr.res.in/~jroy/GC.html}}. The first discovery came from the globular cluster NGC~6093. Using the DM predicted by the YMW16 electron density model and the DM search plan defined in Das et al. 2025~\cite{Das_2025}, we discovered J1617$-$2258A, a 4.32 ms MSP with a DM of 66.8 pc/cc. Subsequent timing follow-up revealed that it is in a highly eccentric (e $\sim$ 0.54) binary orbit with an orbital period of $\sim$ 19 hours and a low-mass companion of $\rm \sim\:0.082\:M_{\odot}$. The large eccentricity and compact orbit enabled the measurement of a post-Keplerian parameter, the periastron advance, which was found to be about 0.5 degrees/yr for this system. Further details are provided in Das et al. 2025~\cite{Das_2025}.

The second discovery was made in NGC~1904. Following a similar approach using the predicted DM of the YMW16 model, we detected J0524$-$2431A, a 4.68 ms pulsar with a DM of 62.3 pc/cc.  Although the detection is relatively faint, follow-up observations have confirmed that it is a binary MSP in a compact orbit with an orbital period of $\sim$ few days.

\begin{figure}
    \centering
    \includegraphics[width=\linewidth]{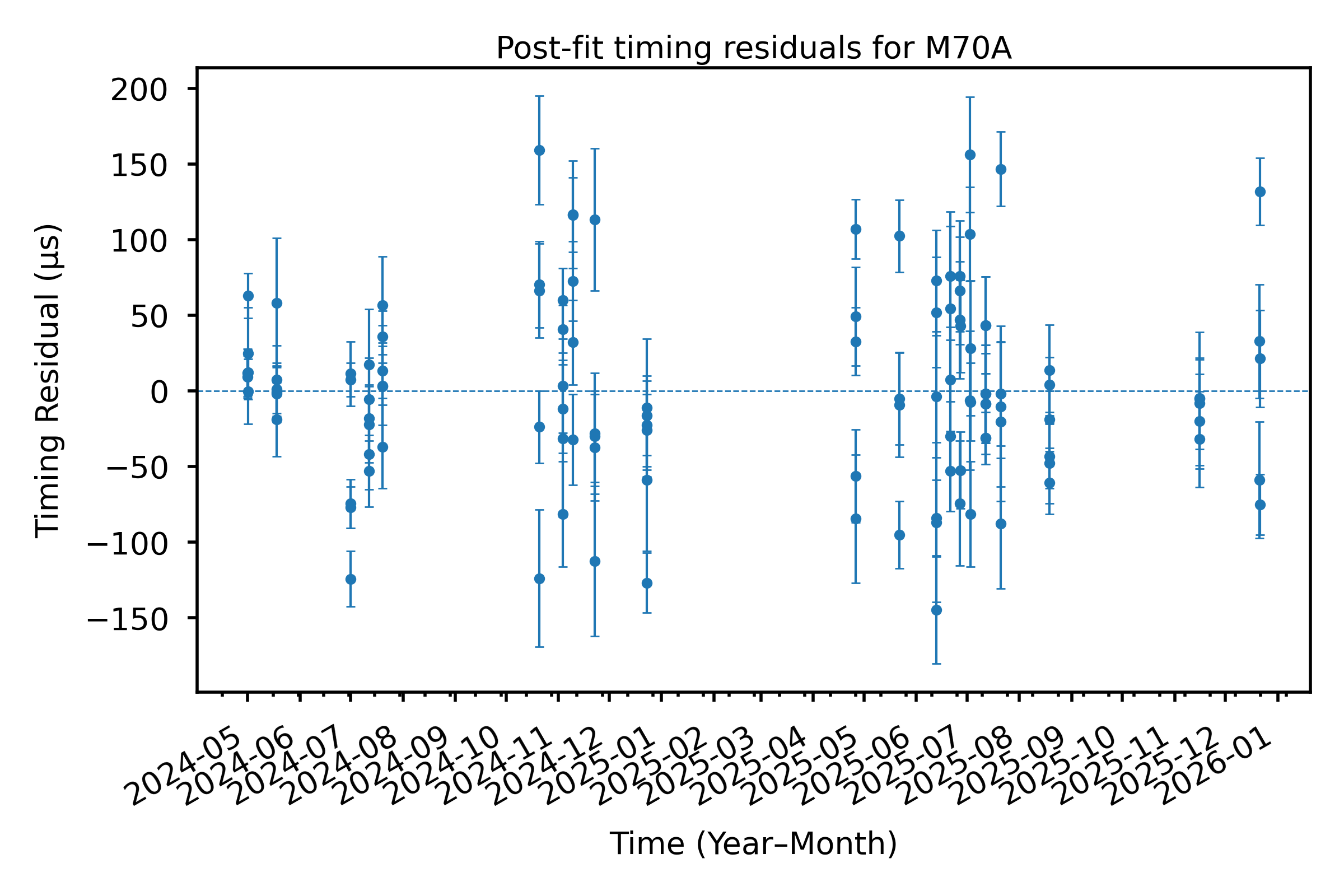}
    \caption{Post-fit timing residual plot of the unique timing solution for the first MSP (J1843-3217A) in NGC 6681 (M70).}
    \label{fig:M70A_timing_residual}
\end{figure}

The remaining five discoveries were from the year 2024, comprising two MSPs in NGC~6637 and three in NGC~6681 (see Das et al. 2026a~\cite{Das_2026a}). Follow-up timing and targeted observations enabled us to localise the first MSP of NGC~6637 (J1831$-$3220A) in the image plane (see Figure~\ref{fig:NGC6637A_localisation}) and to derive unique timing solutions for three out of the five MSPs. In NGC~6637, J1831$-$3220A (Ps = 3.80 ms, DM = 82.12 pc/cc) was found to be a binary MSP in a $\sim$ 1.55-day orbit with a minimum companion mass of $\rm \sim\: 0.18\:M_{\odot}$. In NGC~6681, J1843$-$3217A (Ps = 3.92 ms, DM = 70.6 pc/cc) and J1843$-$3217B (Ps = 4.44 ms, DM = 71.1 pc/cc) were both identified as isolated MSPs (for M70A, see Figure~\ref{fig:M70A_timing_residual}). For the remaining two MSPs, J1831$-$3220B (Ps = 4.80 ms, DM = 81.92 pc/cc) and J1843$-$3217C (Ps = 6.06 ms, DM = 70.1 pc/cc), phase-coherent timing solutions are still under development.

Alongside the discovery, localisation and timing analysis, we also imaged all GCGPS-observed GC fields for non-detection studies. The resulting images are publicly available~\cite{Das_img_2025} to facilitate image-based targeted searches and follow-up studies for GC MSPs.

\subsection{GCGPS PHASE II overview}
\begin{figure}
    \centering
    \includegraphics[width=\linewidth]{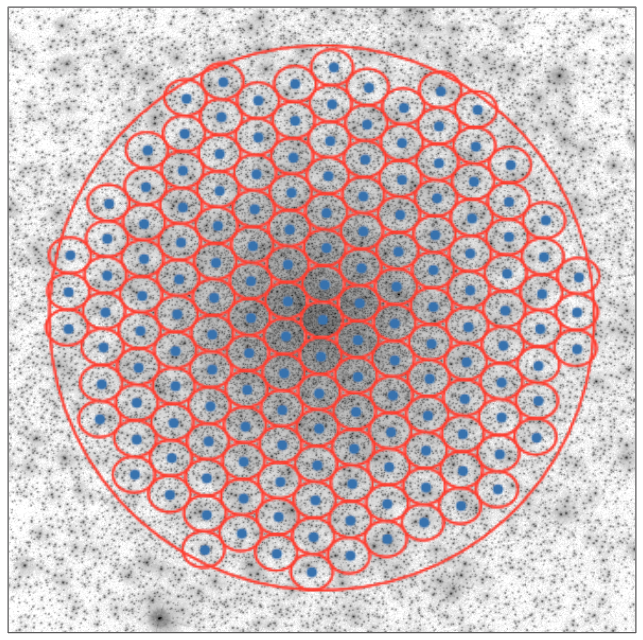}
    \caption{Tiling of the 160 SPOTLIGHT phased-array beams on NGC~6093 (M80) as part of the GCGPS Phase~II survey. The smaller red circles are the half-power line of each post-correlation (PC) beam, while the outer red circle indicates the total field-of-view (FoV) of the total 160 PC beams. The background optical image is taken from the {\tt Hubble UV GC  survey (HUGS)}.}
    \label{fig:SPOTLIGHT_tiling}
\end{figure}

Following the completion of Phase I search observations, we initiated Phase II of the GCGPS survey. In Phase I, the use of only two simultaneous PA beams resulted in our limited sky coverage, and the intrinsic beam response resulted in reduced sensitivity toward the outer regions of the globular clusters. To overcome this limitation, we initiated Phase II of the GCGPS survey. In Phase II, we decided to form a total of 160 post-correlation (PC, see Roy et al. 2018~\cite{Roy_2018}) beams in a compact tiled pattern (see Figure~\ref{fig:SPOTLIGHT_tiling}), enabling uniform sensitivity coverage across the entire cluster field and mitigating the sensitivity loss associated with single-pointed beam observations. This beam formation and data recording were made possible by the newly commissioned SPOTLIGHT\footnote{The SPOTLIGHT website: \url{https://spotlight.ncra.tifr.res.in/}} backend at the uGMRT. Further details about the SPOTLIGHT project can be found in Roy et al. 2024~\cite{Roy_2024}.

The inclusion of 160 PC beams has significantly enhanced both sensitivity and sky coverage. Furthermore, the implementation of the in-field phasing technique by Kudale et al. 2024~\cite{Kudale_2024}, which mitigates ionospheric phase perturbations by enabling iterative phasing of the array directly on the target source throughout the observing duration, eliminates the need for interleaved phase-calibrator scans. Thus, with the 160$-$beam SPOTLIGHT system combined with in-field phasing, the full array is utilised to form each PC beam, substantially enhancing the per-beam sensitivity (by at least a factor of $\sim$ 1.5) while simultaneously providing complete coverage of the GC region. Unlike Phase I, which was constrained by sensitivity and limited sky coverage, Phase II observations are expected to surpass the theoretical sensitivity limits of the TRAPUM survey by a factor of $\sim$ 2, along with exceptional sky coverage, which allowed us to include nearly all southern-hemisphere GCs accessible to the uGMRT sky in GCGPS Phase II.

To process this large-volume multi-beam dataset, we have developed a multi-GPU pulsar search pipeline {\tt ver0} (see Das et al. 2026b~\cite{Das_2026b}) based on {\tt Astro-Accelerate}\footnote{AA GitHub repository: \url{https://github.com/AstroAccelerateOrg/astro-accelerate}}. This pipeline facilitates a GPU-based multi-beam pulsar search capability, with a significant speedup over the legacy GCGPS pipeline {\tt PSS}. This pipeline {\tt ver0} has been implemented as part of the SPOTLIGHT pulsar search project\footnote{{\tt ver0} GitHub repository: \url{https://github.com/jyotirmoydas5392/spotlight_pulsar_search_pipeline.git}}.

This pipeline ({\tt ver0}) has also been deployed as an observatory-level pipeline, supporting the pulsar search survey for open-sky observations with the SPOTLIGHT system. Scientific results from the GCGPS Phase II survey will be reported in upcoming publications.

\section{Summary}
This paper summarises the GCGPS survey, from its initial design and execution to the conclusion of Phase I, along with the implementation of its Phase II component. From GCGPS Phase I, we report the discovery of seven MSPs across four globular clusters. Several of these MSPs were localised using imaging techniques, followed by targeted beamformed timing observations (see Figure~\ref{fig:NGC6637A_localisation}). Timing follow-up with the uGMRT has revealed that three out of the seven MSPs are in binary systems, two are isolated, and two remain to be fully characterised. With the transition to SPOTLIGHT-enabled Phase II, GCGPS marks a step change in sensitivity, completeness, and discovery potential for GC MSPs. In addition to 160 PC beams, we employ the in-field phasing technique to maintain optimal coherence and mitigate sensitivity loss in the highly sensitive PC beams over long observing durations. Data processing for GCGPS Phase II is currently ongoing using the newly developed multi-GPU search pipeline {\tt ver0}. GCGPS thus positions the GMRT as a leading facility for systematic millisecond pulsar discovery and characterisation in globular clusters ahead of the Square Kilometre Array (SKA).

\section{Acknowledgements}
We acknowledge the support of the Department of Atomic Energy, Government of India, under project No. $\rm 12-R\&D-TFR5.02-0700$. The GMRT is run by the National Centre for Radio Astrophysics (NCRA) of the Tata Institute of Fundamental Research, India. We acknowledge the contributions of the SPOTLIGHT team in the development of the multi-GPU pulsar search pipeline. SPOTLIGHT at the GMRT enables studies of pulsars and fast radio bursts (FRBs) using a petaflop-scale computing facility (called Param Brahmand) based on Param Rudra servers, funded by the National
Supercomputing Mission (NSM) of the Government of India (GoI). We especially thank our colleagues at C-DAC (Centre for Development of Advanced Computing) for their support in setting up the Param Brahmand data centre at the GMRT. We thank the uGMRT operators for their coordinated effort in conducting the GCGPS Phase I and Phase II survey observations.

%
%
\noindent\small
Jyotirmoy Das is with the National Centre for Radio Astrophysics (NCRA-TIFR), Spicer College Road, Pune, 411007, India; e-mail: tataidas5392@gmail.com\\[1ex]
Jayanta Roy is with the National Centre for Radio Astrophysics (NCRA-TIFR), Spicer College Road, Pune, 411007, India; e-mail: jroy@ncra.tifr.res.in\\[1ex]
Paulo C. C. Freire is with the Max-Planck-Institut für Radioastronomie (MPIfR),auf dem Hügel 69, Bonn, 53121, Germany; e-mail: pfreire@mpifr-bonn.mpg.de\\[1ex]
Scott Ransom is with the National Radio Astronomy Observatory (NRAO), Charlottesville, Virginia, 22903, United States; e-mail: sransom@nrao.edu\\[1ex]
Bhaswati Bhattacharyya is with the National Centre for Radio Astrophysics (NCRA-TIFR), Spicer College Road, Pune, 411007, India; e-mail: bhaswati@ncra.tifr.res.in\\[1ex]
Karel Adamek is with the Institute of Physics in Opava, Silesian University in Opava, nám. Bezručovo 1150, 746 01 Opava 1-Předměstí, Czechia; e-mail: karel.adamek@gmail.com\\[1ex]
Wesly Armour is with the Oxford e-Research Centre (OeRC), University of Oxford, 7 Keble Rd, Oxford, OX1 3QG, United Kingdom; e-mail: wes.armour@oerc.ox.ac.uk\\[1ex]
Sanjay Kudale is with the National Centre for Radio Astrophysics (NCRA-TIFR), Spicer College Road, Pune, 411007, India; and also with the Giant Metrewave Radio Telescope (GMRT), Khodad, Pune, 410504, India; e-mail: kudale.sanjay@gmail.com\\[1ex]
Mekhala Muley is with the Giant Metrewave Radio Telescope (GMRT), Khodad, Pune, 410504, India; e-mail: mekhala.gmrt@gmail.com
\end{document}